\documentclass[pra,showpacs,amsfonts,amssymb,amsmath]{revtex4}
\usepackage{times}
\usepackage{graphicx}

\begin{document}

\title{Stability analysis for $n$-component Bose-Einstein condensate}

\author{David C. Roberts$^1$ and Masahito Ueda$^{2,3}$} 
\affiliation{$^1$Laboratoire de Physique Statistique de l'Ecole Normale Sup\'erieure, Paris, France \\ 
$^2$Department of Physics, Tokyo Institute of Technology, Tokyo 152-8551, Japan\\ 
$^3$ERATO, Japan Science and Technology Corporation (JST), Saitama 332-0012, Japan} 
\begin{abstract}
We derive the dynamic and thermodynamic stability conditions for dilute multicomponent Bose-Einstein condensates (BECs).  These stability conditions, generalized for $n$-component BECs, are found to be equivalent and are shown to be consistent with the phase diagrams of two- and three-component condensates that are derived from energetic arguments.  
\end{abstract}

\maketitle

\section{Introduction}

Interpenetrating superfluids have intrigued physicists for decades.  The possibility of a superfluid mixture of two different isotopes of Helium motivated many early theoretical studies, including a hydrodynamic approach \cite{Khalatnikov} and an approach using the Bogoliubov analysis \cite{colson}. (See also \cite{bachius} for a generalization of the Bogoliubov approach to mixtures in a different context.) However, these mixtures have never been experimentally realized in superfluid Helium. 

Recent advances in creating and manipulating Bose-Einstein condensates in ultracold alkali gases have made it possible to probe the properties of interpenetrating superfluids by studying, for instance, two internal states of a single atom \cite{jila}, different atomic species \cite{diffatom,ospelkaus}, different isotopes of a single atom, or a combination thereof.  One such advancement is the achievement of a condensate of Ytterbium \cite{yt} which, with its five naturally occurring bosonic isotopes, opens the door for the creation of interpenetrating condensates with up to five components.  Another is the development of optical Feshbach resonance techniques \cite{theis}.  Unlike magnetic Feshbach resonance techniques, where typically only one scattering length can be tuned at a time, these new optical methods would, in principle, allow one to tune both the interspecies and intraspecies scattering lengths to probe a much more significant portion of parameter space \cite{tak}.  

In this paper, we investigate the fundamental stability and phase diagrams of such multicomponent dilute condensates, excluding effects such as external potentials and metastability.  We begin by deriving a simple relation to determine the thermodynamic and the dynamic stability conditions of an $n$-component miscible mixture, and we show that the stability conditions, at least within our approximations, are equivalent.  Then, through an energetic approach, we discuss these phase diagrams of two-component mixtures and those of three-component mixtures and demonstrate that these are consistent with the stability conditions. 

\section{Stability conditions}
Let us consider a general miscible $n$-component condensate mixture in a box of volume $V$ with periodic boundary conditions \cite{per} in the thermodynamic limit described by the following Hamiltonian
\begin{equation}
\hat H = \sum^{n}_{i=1} \int d^3 {\bf r} \left[ -\hat \psi^{\dag}_i ({\bf r} ) \frac{ \hbar^2 \vec{\nabla}^2}{2 m} \hat \psi_i({\bf r})  +\frac{g_{ii}}{2} \hat \psi^{\dag}_i({\bf r}) \hat \psi^{\dag}_i({\bf r}) \hat \psi_i({\bf r}) \hat \psi_i({\bf r}) + \sum_{j (\ne i)} g_{ij} \hat \psi^{\dag}_i({\bf r}) \hat \psi^{\dag}_j({\bf r}) \hat \psi_i({\bf r})\hat \psi_j({\bf r}) \right]
\end{equation}
where, for simplicity, we assume that all species of atoms have the same mass $m$.  We also assume that the number of particles of each species is fixed and cannot be converted into a different species.  The atomic interactions are characterized by the interparticle contact pseudopotential where the coupling constants, $g_{ij}=4 \pi \hbar^2a_{ij}/m$, can be given in terms of the two-body s-wave scattering length $a_{ij}$.  

Let us expand the bosonic field operators in terms of plane waves, $\hat \psi_i({\bf r}) = \sum_{\bf k} \hat a^i_{\bf k} e^{i {\bf k} \cdot {\bf r}}/\sqrt{V}$, where $\hat a^i_{\bf k}$ annihilates a particle of species $i$ in state $\bf k$ and obeys the usual bosonic commutation relations.  Let us assume that each species is condensed such that a macroscopic number of atoms are in state ${\bf k}=0$.   Following the standard Bogoliubov approximation \cite{Bog}, we assume $\hat a^i_0$ and $\hat a^{i \dag}_0$ to be commuting c-numbers equal to $\sqrt{N^i_0}$, where $N^i_0$ is the number of particles in the condensate of species $i$.  Assuming that $N^i-N^i_0 \ll N^i$, where $N^i=N^i_0+ \sum_{{\bf k} \ne 0} \hat a_{\bf k}^{i \dag} \hat a^i_{\bf k}$ is the total number of particles in species $i$ (alternatively one could introduce a chemical potential in the grand canonical ensemble with $N^i$ given by miminimizing the thermodynamic potential), we can approximate the Hamiltonian to leading order as
\begin{equation}
\hat H_0 = \frac{1}{2} \int d^3 {\bf r} \sum_{i=1}^n \left[ g_{ii} \rho_i^2+\sum_{j (\ne i)} 2 g_{ij} \rho_i \rho_j \right],
\end{equation}
where $\rho_i=N_0^i/V$.  Note that the kinetic energy plays a negligible role because we consider only atoms in the $k=0$ state.  The energy density of an $n$-component miscible mixture in the mean field approximation can therefore be written as
\begin{equation}
{\cal E}^{\rm mi}_n = \frac{1}{2} \rho_i \rho_j g_{ij}.
\end{equation}
In this homogenous system, thermodynamic stability implies that the Hessian matrix defined by $\partial_{\rho_i} \partial_{\rho_j} {\cal E}^{\rm mi}_n$ is positive semi-definite \cite{thermo}, i.e. all eigenvalues of the matrix are non-negative, which is still possible in certain parameter regimes even if the interspecies scattering lengths are negative. Because $\partial_{\rho_i} \partial_{\rho_j} {\cal E}^{\rm mi}_n=g_{ij}$, thermodynamic stability of this $n$-component miscible mixture implies that the matrix $g_{ij}$ must be positive semi-definite.

Now let us consider the dynamic stability of the system.  Specifically, we investigate the conditions under which a small density perturbation increases exponentially in time or, equivalently, the excitation spectrum becomes imaginary.  We derive the excitation dispersion relation by expanding the Hamiltonian of the system to the next order in the Bogoliubov approximation described above to arrive at
\begin{equation}
\hat H \approx \hat H_0 + \hat H_{\rm intra} + \hat H_{\rm inter}
\end{equation}
where the intraspecies contribution to the Bogoliubov Hamiltonian is given by 
\begin{equation}
\hat H_{\rm intra} =\sum^N_{i=1} \sum_{{\bf k} \ne 0} \frac{1}{2} \left[2 (t^0_k+\rho_i g_{ii})( \hat a_{\bf k}^{i \dag} \hat a^i_{\bf k})+\rho_i g_{ii} ( \hat a_{\bf k}^{i \dag} \hat a^{i \dag}_{-{\bf k}} +\hat a_{\bf k}^{i} \hat a^i_{\bf {-k}}) \right]
\end{equation}
and $t^0_k=(\hbar k)^2/2m$. The interspecies contribution is given by
\begin{equation}
\hat H_{\rm inter}=\sum^n_{i=1} \sum_{j (\ne i)} \sum_{{\bf k} \ne 0} \frac{\sqrt{N^i N^j} g_{ij}}{V} \left[ ( \hat a_{\bf k}^{j \dag} \hat a^i_{\bf k} +\hat a_{\bf {k}}^{i \dag} \hat a^j_{\bf {k}})+( \hat a_{\bf k}^{j \dag} \hat a^{i \dag}_{-\bf k} +\hat a_{\bf {k}}^{j} \hat a^i_{\bf {-k}}) \right].
\end{equation}
In this approximation, the Hamiltonian is quadratic and, as such, can be diagonalized by canonical transformation.    Therefore, one can transform this Hamiltonian describing weakly interacting particles into a Hamiltonian describing non-interacting quasiparticles with $n$ independent modes of oscillation.  The excitation energies $\hbar \omega_n(k)$ can be derived by assuming a harmonic time dependence $e^{-i \omega_n(k) t}$ of the particle operators. Writing the Heisenberg equations of motion for the particle annihilation and creation operators, we obtain
\begin{equation}
\hbar \omega_n(k) \hat a_{\bf k}^i = (t^0_k+g_{ii} \rho_i) \hat a_{\bf k}^i + g_{ii} \rho_i \hat a^{i \dag}_{-\bf k} +\sum^n_{j \ne i} \frac{\sqrt{N^i N^j} g_{ij}}{V} (\hat a_{\bf k}^j+a^{j \dag}_{-\bf k})
\end{equation}
\begin{equation}
\hbar \omega_n(k) a^{i \dag}_{-\bf k} = -(t^0_k+g_{ii} \rho_i) \hat a^{i \dag}_{-\bf k}  - g_{ii} \rho_i \hat a_{\bf k}^i -\sum^n_{j \ne i}\frac{\sqrt{N^i N^j} g_{ij}}{V} (\hat a_{\bf k}^j+\hat a^{j \dag}_{-\bf k}).
\end{equation}
The excitation energies are thus given by the eigenvalues of the $2n$ by $2n$ matrix given by the above equations and are given by
\begin{equation}
[\hbar \omega_n(k)]^2 = t^0_k(t^0_k+2 \lambda_n )
\end{equation}
where $\hbar \omega_n(k)$ is the excitation energy and $\lambda_n$ are the eigenvalues of the matrix $\Lambda_{ij}=\rho_j g_{ij}$.  If any of these eigenvalues are negative then the excitation energy becomes imaginary and the system becomes dynamically unstable in the sense that infinitesimal perturbations will grow exponentially with time.  The dynamic instability starts at low momenta, and thus the condition for the system to remain dynamically stable is that the matrix $\Lambda_{ij}$ be positive semi-definite.  Since $\rho_j$ are positive and positive semi-definiteness implies that all upper left submatrices are positive, the condition of $\Lambda_{ij}$ being positive semi-definite is equivalent to the condition of $g_{ij}$ being positive semi-definite.   

The fact that the dynamic stability condition, which indicates whether a state is metastable or not, and the thermodynamic stability condition are the same implies that the energy density function's local and global extrema are the same.  The energy surface is therefore downward convex for stable systems and upward convex for unstable systems \cite{hess}.  The stability condition of this system, namely that $g_{ij}$ must be positive semi-definite, has another important consequence.  If the condition is met then the determinant of all upper left submatrices are positive implying that, for an $n$-component miscible system to be stable, all subsets of the species comprising the system must also independently satisfy the relevant stability conditions. A restrictive hierarchy of stability conditions is thus defined.

\section{Phase diagrams}
To illustrate these stability conditions, we will now discuss the ultimate fate of these mixtures when they are unstable; we consider one-, two- and three-component systems in that order.  We shall rely on purely energetic arguments in the mean-field approximation.  We shall make the assumption that the system has been allowed a sufficiently long time to relax to its lowest energy state and ignore metastable states such as molecules.  We ignore all effects from the thickness of the interface between regions and surface tension \cite{ao} as we assume that the kinetic energy plays a negligible role.  We also assume that the coupling constants remain unchanged in the presence of other atoms whose effects we briefly discuss below.  

For the one-component case, the stability conditions yield the well known result that $g_{11}>0$ must be positive in order for the condensate to be stable.  If $g_{11}<0$ then one can formally prove that the system will collapse in density space in a finite time if $H(t=0)<0$ (where $H$ is the full classical Hamiltonian) and in the thermodynamics limit \cite{sulem}.  In realistic systems, however, other effects such as three-body recombination will become important \cite{saito}.

For a system of two components, in this simplified mean field approximation we can safely make the assumption that only two stable states exist:  the miscible case given by 
\begin{equation}
{\cal E}_2^{\rm mi}=\frac{g_{11} \rho_1^2}{2}+\frac{g_{22} \rho_2^2}{2} +g_{12} \rho_1 \rho_2,
\end{equation}
and the completely immiscible case given by
\begin{equation}
\label{twoim}
{\cal E}_2^{{\rm im}}=\frac{g_{11} \rho_1^2}{2}+\frac{g_{22} \rho_2^2}{2} +\sqrt{g_{11} g_{22}} \rho_1 \rho_2.
\end{equation}
This last equation can be easily derived if we assume that the total energy of the immiscible case is given by
\begin{equation}
E_2^{{\rm im}}=\frac{g_{11} N_1^2}{2 V_A}+\frac{g_{22} N_2^2}{2 V_B} 
\end{equation}
where $V_A$ and $V_B$ are the volumes of regions where species 1 and species 2 exist, respectively.    We can simplify $E^{{\rm im}}_2$ by noting that the total volume is fixed, i.e. $V_A+V_B=V$, and, by either assuming that the system is mechanically stable \cite{stringari}, i.e. $\partial E_2^{{\rm im}}/ \partial V_A = \partial E_2^{{\rm im}}/ \partial V_B$, or, equivalently, minimizing the energy with respect to $V_A$ under the constraint of $V_A+V_B=V$ \cite{ao}, we arrive at eq. (\ref{twoim}).

It is clear that when $g_{12}^2>g_{11}g_{22}$ and $g_{12}$ is positive the system would energetically prefer to be in the immiscible phase.  This condition is consistent with both stability conditions of the miscible phase derived above which indicate that if  $g_{12}^2>g_{11}g_{22}$ then the miscible system is unstable.  If $g_{12}^2>g_{11}g_{22}$ and $g_{12}$ is negative then we would expect the system to collapse in a finite time because it can be shown that if $H(t=0)<0$ for a mixture, the system will exhibit finite time collapse \cite{robertsnewell}.  

\begin{figure}[]
\includegraphics[scale=0.7]{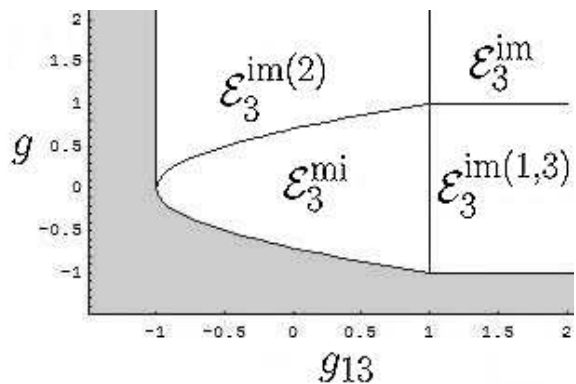}
\caption{\label{phase} 
The phase diagram for a three-component system is plotted against interspecies coupling constants $g_{12}=g_{23}=g$ and $g_{13}$ under the assumptions of $N_1=N_2=N_3$ and $g_{11}=g_{22}=g_{33}=1$.  Note that the shaded region is expected to collapse in density space if $H(t=0)<0$.}
\end{figure}

We now turn to the rich phase diagram of systems with three interacting species.  For simplicity we will assume that all species have an equal number of particles, i.e. $N_1=N_2=N_3$, leading to a uniform constant particle density $\rho$.  We also assume that the intraspecies coupling constants are normalized by the interspecies coupling constants and are equal, i.e. $g_{11}=g_{22}=g_{33}=1$.  Here, the possible stable states are the miscible phase where energy density is given by
\begin{equation}
{\cal E}_3^{\rm mi}/\rho^2=\frac{3+2(g_{12}+g_{23}+g_{13})}{2};
\end{equation}
 the completely immiscible phase where each species is separated from the other such that
\begin{equation}
{\cal E}_3^{{\rm im}}/\rho^2=\frac{9}{2};
\end{equation}
a phase in which species $i$ is separated from a region where species $j$ and $k$ are miscible, with an energy density given by
\begin{equation}
{\cal E}_3^{{\rm im}(i)}/\rho^2=\frac{3+2 g_{jk}+2 \sqrt{2} \sqrt{1+g_{jk}}}{2};
\end{equation}
and a phase where species $i$ and $j$ are miscible, $i$ and $k$ are miscible, but $j$ and $k$ are immiscible, with an energy density per atom given by
\begin{equation}
{\cal E}_3^{{\rm im}(j,k)}/\rho^2=\frac{3+2g_{ij}+2 g_{ik}+\frac{g_{ij}}{g_{ik}}+\frac{g_{ik}}{g_{ij}}}{2}.
\end{equation}
As in the earlier two-species example, we have assumed the pressures in each separate region to be equal, that is, we have assumed mechanical equilibrium.  In the stable state, ${\cal E}_3^{{\rm im}(j,k)}$, we have also assumed the chemical potentials of species $i$ in the different regions to be the same.   The phase diagram for the three-component case is shown in Fig. \ref{phase} where we have assumed $g_{12}=g_{23}=g$ for simplicity.  In this case, ${\cal E}_3^{{\rm im}(1,2)}$ and ${\cal E}_3^{{\rm im}(2,3)}$ are stable only when $g_{13}=g$. The shaded region in the figure will collapse in density space if $H(t=0)<0$ \cite{robertsnewell}. 

\section{Discussion}
Throughout this mean field analysis, we have assumed that coupling constants are determined entirely by two-body physics and remain unchanged in the presence of condensates.  If many-body effects are taken into account, i.e. that the coupling constants may be modified in the presence of a foreign species, many exotic phases could occur, such as the possibility of a three-component condensate having a stable state with three distinct regions, each with different pairs of miscible species.  In addition, other interesting effects would be possible, such as the presence of a third species making two otherwise immiscible species miscible, or the destabilization of a previously stable two species miscible mixture by the introduction of a third species.  Finally, effects not considered in this paper, such as metastable states \cite{miesner} and a non-uniform trapping potential \cite{ho}, enrich the picture even further and come into play in realistic experiments involving trapped dilute condensates.  

DCR gratefully acknowledges many enlightening talks with Yves Pomeau, Alexander Fetter, Jean-No\"el Fuchs and Yoshiro Takahashi and  as well as the and the generous hospitality of the Departments of Physics at Kyoto University and the Tokyo Institute of Technology. The authors would also like to acknowledge the support of the Institute of Nuclear Theory at the University of Washington where part of this work was carried out.


\begin{thebibliography}{9}

\bibitem{Khalatnikov} Khalatnikov. JETP-USSR {\bf 5}, 542-545 (1957)

\bibitem{colson} W. B. Colson and A. L. Fetter.  J. Low Temp. Phys.  {\bf 33}, 231 (1978)

\bibitem{bachius} W. H. Bassichis.  Phys. Rev.  {\bf 134}, A543 (1964)

\bibitem{jila} C. J. Myatt, E. A. Burt, R. W. Ghrist, E. A. Cornell and C. E. Wieman.  Phys. Rev. Lett. {\bf 78}, 586 (1997)

\bibitem{diffatom} G. Modugno, M. Modugno, F. Riboli, G. Roati, and M. Inguscio.  Phys. Rev. Lett. {\bf 89}, 190404 (2002)

\bibitem{ospelkaus} C. Ospelkaus, S. Ospelkaus, K. Sengstock, and K. Bongs.  Phys. Rev. Lett. 96, 020401 (2006)


\bibitem{yt}  Y. Takasu, K. Maki, K. Komori, T. Takano, K. Honda, M. Kumakura, T. Yabuzaki, and Y. Takahashi. Phys. Rev. Lett. {\bf 91} 040404 (2003)  

\bibitem{theis} M. Theis, G. Thalhammer, K. Winkler, M. Hellwig, G. Ruff, R. Grimm, and J. Hecker Denschlag.  Phys. Rev. Lett. {\bf 93}, 123001 (2004)

\bibitem{tak} Yoshiro Takahashi, Private communication.

\bibitem{per} In this letter, we exclude the possibility that an unstable system can remain metastable because of the finite period. 

\bibitem{Bog} N. Bogoliubov. J. Phys. (U.S.S.R) {\bf 11}, 23 (1947)

\bibitem{thermo} Thermodynamic stability implies that, at a critical point in density-space, there is an unconstrained local minimum.  In this case, however, the minimum densities are determined by the assumption that the system comprises a fixed number of particles of each species in a rigid container of volume V; the critical point thus coincides with the minimum densities so constrained.

\bibitem{hess} This can also be seen from the behavior of the Hessian matrix of the energy density: If the matrix is independent of the densities $\rho_i$, which means that if it is positive semi-definite for one point it is semi-positive definite for all points, then the energy surface is convex. 

\bibitem{sulem} C. Sulem and P. L. Sulem, The nonlinear Schrodinger equation: self-focusing
and wave collapse. (Springer, 1999).

\bibitem{saito} H. Saito and M. Ueda,  Phys. Rev. Lett. 90, 040403 (2003)

\bibitem{stringari} L. Pitaevskii and S. Stringari.  "Bose-Einstein Condensation".  Clarendon Press, Oxford (2003) 

\bibitem{ao} P. Ao and S. T. Chui.  Phys. Rev. A {\bf 58}, 4836 (1998)

\bibitem{robertsnewell} D. C. Roberts and A. C. Newell.  cond-mat/0605603

\bibitem{miesner} H.-J. Miesner, D. M. Stamper-Kurn, J. Stenger, S. Inouye, A. P. Chikkatur, 
and W. Ketterle.  Phys. Rev. Lett. 82, 2228 (1999)

\bibitem{ho} T. L. Ho and V. B. Shenoy.  Phys. Rev. Lett. {\bf 77}, 3276 (1996)

\end{thebibliography}
\end{document}